\documentstyle[preprint,psfig,aps]{revtex}
\begin{document}
\bibliographystyle{plain}

\title{ 
Ladders in a magnetic field: a strong coupling approach
}
\vskip0.5truecm 
\author{Fr\'ed\'eric Mila}

\address{Laboratoire de Physique 
Quantique, Universit\'e Paul Sabatier, 118 Route de Narbonne, 31062 
Toulouse Cedex, France. }\vskip0.5truecm
      
\maketitle

\begin{abstract}
\begin{center} 
\parbox{14cm}{We show that non-frustrated and
frustrated ladders in a magnetic field 
can be systematically mapped onto an XXZ Heisenberg model in a longitudinal
magnetic field 
in the limit where the rung coupling is
the dominant one. This mapping is valid in the critical region
where the magnetization goes from zero to saturation. It allows one to relate
the properties of the critical phase ($H_c^1$, $H_c^2$, the critical exponents)
to the exchange integrals and provide quantitative estimates of the frustration
needed to create a plateau at half the saturation value for different models of
frustration.
}
\end{center}
\end{abstract}
\vskip .1truein
 
\noindent PACS Nos : 75.10.jm 75.40.Cx 75.50.Ee
\vskip2pc
\newpage

\section{\bf Introduction}

Intermediate between 1D and 2D, ladders have been the subject of an impressive
amount of work over the past few years\cite{dagotto}.
Thanks to an intensive
experimental\cite{hammar,chaboussant1,chaboussant2,chaboussant3}
and theoretical\cite{hayward,chitra,weihong} effort, quite a lot is
understood concerning the properties of S=1/2 ladders in a magnetic field. In
particular, the magnetization starts to increase above a magnetic field $H_c^1$
and saturates above a magnetic field $H_c^2$, and the phase realized for
intermediate magnetic fields is believed to be a Luttinger liquid with gapless
excitations and a power law decay of the correlation functions. As usual, it is
difficult however starting from a microscopic description in terms of exchange
integrals to calculate the parameters of the low energy theory in the Luttinger
liquid phase, and this description is to a certain extent phenomenological.

In this paper, we show explicitly how this problem can be mapped onto 
the XXZ model in a longitudinal magnetic 
field if the rung coupling is the dominant one. This is actually the case of 
Cu$_2$(C$_5$H$_{12}$N$_2$)$_2$Cl$_4$, the ladder system on which most results
under strong magnetic fields have been obtained so 
far\cite{hammar,chaboussant1,chaboussant2,chaboussant3}.
Such a mapping can actually be
performed for any type of coupling between the rungs, and we will 
study the following Hamiltonian (see Fig. 1)

\begin{eqnarray}
{\cal H} & = & J_\perp \sum_{i=1}^N \vec S_{i,1} \vec S_{i,2}
+J_1 \sum_{i=1}^N \sum_{\alpha=1}^2 \vec S_{i,\alpha} \vec S_{i+1,\alpha} 
+J_2' \sum_{i=1}^N \vec S_{i,1} \vec S_{i+1,2} \nonumber \\
& + & J_2'' \sum_{i=1}^N \vec S_{i,2} \vec S_{i+1,1}
-H \sum_{i=1}^N \sum_{\alpha=1}^2 S_{i,\alpha}^z
\label{hamiltonian}
\end{eqnarray}

In this expression, $\alpha$ (resp. $i$) is a chain (resp. rung) 
index, $N$ is the total number of rungs, and periodic boundary conditions along
the chain direction are implicit. If all the couplings except $J_\perp$ 
are set to zero, the system is a
collection of independent rungs. The states of a given rung are denoted by
$|S>=(|\uparrow \downarrow> - | \downarrow \uparrow>)/\sqrt{2}$,
$|T_1>=|\uparrow \uparrow >$, 
$|T_0>=(|\uparrow \downarrow> + | \downarrow \uparrow>)/\sqrt{2}$ and
$|T_{-1}>=|\downarrow \downarrow>$. In a configuration $|\sigma_1 \sigma_2>$,
$\sigma_1$ (resp. $\sigma_2$) refers to chain 1 (resp. 2). Their energies are
$E(S)=-3J_\perp/4$, $E(T_1)=J_\perp/4-H$, $E(T_0)=J_\perp/4$ and 
$E(T_{-1})=J_\perp/4+H$. So upon increasing the magnetic field 
the groundstate of a given rung undergoes a transition between the 
singlet $|S>$ and the triplet 
$|T_1>$ at $H_c=J_\perp$, and the total magnetization of
the system jumps discontinuously from zero to saturation.

If the other couplings are non-zero but small, this abrupt transition is
expected to broaden between $H_c^1$ and $H_c^2$, $H_c^2-H_c^1$ being of 
the order of the largest of the couplings $J_1$, $J_2'$ and $J_2''$. 
In this limit, the properties 
of the system for $H_c^1 \le H \le H_c^2$ are best understood by splitting the
Hamiltonian into two parts:

\begin{eqnarray}
{\cal H} & = & {\cal H}_0 + {\cal H}_1,  \nonumber \\
{\cal H}_0 & = & J_\perp \sum_{i=1}^N \vec S_{i,1} \vec S_{i,2}
 - H_c \sum_{i=1}^N \sum_{\alpha=1}^2 S_{i,\alpha}^z,  \nonumber\\
{\cal H}_1 & = &  J_1 \sum_{i=1}^N \sum_{\alpha=1}^2 \vec S_{i,\alpha} 
\vec S_{i+1,\alpha} 
+J_2' \sum_{i=1}^N \vec S_{i,1} \vec S_{i+1,2}
+J_2'' \sum_{i=1}^N \vec S_{i,2} \vec S_{i+1,1}
-(H-H_c) \sum_{i=1}^N \sum_{\alpha=1}^2 S_{i,\alpha}^z
\end{eqnarray}

The groundstate of ${\cal H}_0$ is $2^N$ times degenerate since each rung 
can be in the state $|S>$ or $|T_1>$, and the 
first excited state has an energy equal to $J_\perp$. ${\cal H}_1$ will lift the
degeneracy in the groundstate manifold, leading to an effective Hamiltonian
that can be derived by standard perturbation theory.
Let us start by introducing pseudo-spin S=1/2 operators $\vec
\sigma_i$ that act on the states $|S>_i$ and $|T_1>_i$ of rung i according to 

\begin{eqnarray}
\sigma_i^z |S>_i  =  -\frac {1} {2} |S>_i & \ \ \ \ \ & \sigma_i^z |T_1>_i  = 
\frac {1} {2} |T_1>_i \nonumber \\
\sigma_i^+ |S>_i  =  |T_1>_i & \ \ \ \ \ &  \sigma_i^+ |T_1>_i  =  0 
\nonumber \\
\sigma_i^- |S>_i  =  0 & \ \ \ \ \ & \sigma_i^z |T_1>_i  =  |S>_i 
\end{eqnarray}

Then, to first order, and up to a constant, the effective Hamiltonian reads:
\begin{equation}
{\cal H}_{\rm eff}=\sum_{i=1}^N [ J^{\rm eff}_{xy} (\sigma_i^x \sigma_{i+1}^x + 
\sigma_i^y \sigma_{i+1}^y) 
+J^{\rm eff}_z \sigma_i^z \sigma_{i+1}^z ] - H^{\rm eff} \sum_{i=1}^N \sigma_i^z
\label{effective}
\end{equation}
The parameters of 
$H_{\rm eff}$ are
given by
\begin{eqnarray}
J^{\rm eff}_{xy} & = & J_1 - \frac {J_2'} {2} - \frac {J_2''} {2} 
\nonumber \\
J^{\rm eff}_{z} & = & \frac {J_1} {2} + \frac {J_2'} {4} + \frac {J_2''} {4}
\nonumber \\
H^{\rm eff} & = & H-H_c - \frac {J_1} {2} - \frac {J_2'} {4} - \frac {J_2''} {4}
\label{parameters}
\end{eqnarray}

The Hamiltonian of Eq.(\ref{effective}) is nothing but the XXZ model in a
longitudinal magnetic field. This problem has been studied by several authors
over the years, and most of the relevant information concerning the
properties of the model is available in the literature\cite{haldane}. 
In particular, the
exponents of the spin-spin correlation functions have been obtained analytically
when the model is integrable and numerically otherwise. To translate 
these results into the language of the original Hamiltonian of 
Eq.(\ref{hamiltonian}), one just has to 
express the original operators $S_{i,\alpha}^+$, 
$S_{i,\alpha}^-$ and $S_{i,\alpha}^z$ in terms of the pseudo-spin operators.
This can be done by inspection, and the results are:

\begin{eqnarray}
S_{i,1}^+  =  - \frac{1} {\sqrt{2}} \sigma_i^+ & \ \ \ \ \ & 
S_{i,2}^+   =   \frac{1} {\sqrt{2}} \sigma_i^+  \nonumber \\
S_{i,1}^-  =  - \frac{1} {\sqrt{2}} \sigma_i^- & \ \ \ \ \ & 
S_{i,2}^-   =   \frac{1} {\sqrt{2}} \sigma_i^-  \nonumber \\
S_{i,1}^z   =  \frac{1} {2} (\sigma_i^z +  \frac{1} {2}) & \ \ \ \ \ & 
S_{i,2}^z   =  \frac{1} {2} (\sigma_i^z +  \frac{1} {2}) 
\label{translation}
\end{eqnarray}
A detailed discussion of the correlation functions measured in NMR experiments
can be found in Ref.\cite{chitra}.
We now discuss the implications of this 
mapping in
different cases. 

\section{\bf The regular ladder: $J'_2=J_2''=0$}

In that case $J^{\rm eff}_{xy}=J_1$ is twice as large as $J^{\rm eff}_{z}$: We
are in the XY universality class of the anisotropic Heisenberg model. The system
is gapless, and it behaves as a Luttinger liquid. Having explicit expressions of
the coupling constants in terms of the microscopic parameters, one can
calculate everything in terms of these parameters. For instance, we can express
$H_c^1$ and $H_c^2$ in terms of $J_\perp$ and $J_1$. This is most easily done by
first performing a Jordan-Wigner transformation to map the problem onto a
problem of
interacting, spinless fermions:

\begin{equation}
{\cal H}_{SF}=  t \sum_i^N ( c^\dagger_i c_{i+1}^{\vphantom{\dagger}} + {\rm
h.c.} ) + V \sum_i^N n_i n_{i+1} - \mu \sum_i^N n_i
\label{spinless}
\end{equation}

The parameters of this Hamiltonian are given in terms of those of 
Eq.(\ref{effective}) by $t=J^{\rm eff}_{xy}/2$, $V=J^{\rm eff}_{z}$ and 
$\mu = H^{\rm eff}+J^{\rm eff}_{z}$. $H_c^1$ corresponds to the chemical 
potential at which the
band of spinless fermions starts to fill up. In that limit the repulsion term is
irrelevant because the density of spinless fermions vanishes, so that the
chemical potential corresponding to $H_c^1$ is given by $\mu=-2t$. This leads to
the result $H_c^1=J_\perp-J_1$. To estimate $H_c^2$, one cannot neglect the
repulsion term because the band is completely filled. The simplest way to take
it into account is to perform a particle-hole transformation on the Hamiltonian
of Eq.(\ref{spinless}): $c^\dagger_i \rightarrow d_i^{\vphantom{\dagger}}$. Up
to a constant, the new Hamiltonian reads

\begin{equation}
{\cal H}_{\rm hole}= - t \sum_i^N ( d^\dagger_i d_{i+1}^{\vphantom{\dagger}} 
+ {\rm h.c.} ) + V \sum_i^N n_i^d n_{i+1}^d - \mu_h \sum_i^N n_i^d
\label{hole}
\end{equation}
where the hole chemical potential $\mu_h$ is given by $\mu_h = -\mu + 2V$. In
terms of holes, $H_c^2$ corresponds to the chemical potential where the band
starts to fill up,
and one can again neglect the repulsion term. Note however that this is not
equivalent to neglecting the repulsion in Eq.(\ref{spinless}) since $V$ appears
in the expression of $\mu_h$. The chemical potential corresponding to $H_c^2$ is
thus given by $\mu_h=-2t$, leading to $H_c^2=J_\perp + 2 J_1$. These expressions
of $H_c^1$ and $H_c^2$ agree with those of Ref.\cite{chaboussant1} obtained
along different lines, and they compare well with the experimental values
for Cu$_2$(C$_5$H$_{12}$N$_2$)$_2$Cl$_4$\cite{chaboussant1}.  

The same argument actually apply if $J_2'$ and $J_2''$ are not equal to zero.
To first order, $H_c^2$ is unaffected, and $H_c^1$ is given by
$H_c^1=J_\perp-J_1+(J'_2+J''_2)/2$.

\section{\bf The frustrated ladder: $J'_2=J_2''=J_2$}

When $J_2\ne 0$, the effective Hamiltonian is in the
universality class of the XY model only if $J_2$ is not too large: There is a
transition to the Ising universality class when 
$J^{\rm eff}_{xy}=J^{\rm eff}_{z}$, i.e. $J_2=J_1/3$ to first order. In terms of
spinless fermions, the Ising limit means that $V$ is large enough to make the
half-filled system insulating\cite{ovchinikov}. 
The chemical potential as a function of the
band-filling will then have a jump. In the original spin language, 
this implies that
there will be a plateau in the magnetization at half the saturated value as a
function of magnetic field. Note that similar conclusions have been obtained 
along
different lines by several authors\cite{totsuka1,totsuka2,cabra,tonegawa} 
concerning the case $J_2''=0$ (see next section).
In the plateau region, there is an order parameter
corresponding to alternating singlets and triplets on neighbouring rungs. 
For $J_2=J_1$, the effective Hamiltonian becomes purely Ising: 
$J^{\rm eff}_{z}=J_1$, $J^{\rm eff}_{xy}=0$. This result, clearly valid up to
first order after Eq.(\ref{parameters}), is actually exact including all order
corrections. The simplest way to understand this is to realize that a singlet on
a given rung is completely decoupled from the rest because all the exchange 
integrals starting from this singlet belong to a pair of equal exchange 
integrals connecting a spin to both ends of the singlet. So the perturbation
cannot couple to the singlets, and the XY exchange integral must vanish.
Besides, this argument shows that the state with
alternating singlets and triplets is an eigenstate of the Hamiltonian. It
is easy to prove that it is the groundstate for $H$ between $H_c^1$ and 
$H_c^2$, which are given by $H_c^1=J_\perp$ and $H_c^2=J_\perp+2J_1$ in 
the present
case. So the plateau will
extend over all the intermediate region between zero and saturated
magnetization.

\section{\bf The zigzag ladder: $J_2''=0$}

This case is very similar to the previous one. There will be a transition to an
Ising phase when $J'_2$ is large enough, a conclusion already reached by other
authors using different arguments\cite{totsuka1,totsuka2,tonegawa}. 
To first order the critical value is
given by $J'_2=2J_1/3$. The only difference is that there is no pure
Ising phase in that case since a singlet is only decoupled from neighbouring
singlets, and not from neighbouring triplets. The reason for mentioning this
particular case of frustration is that 
it corresponds in principle to the physical situation realized
in Cu$_2$(C$_5$H$_{12}$N$_2$)$_2$Cl$_4$. Our estimate of the critical value of
$J'_2$ to enter the Ising phase can be used as an upper bound to this exchange
integral in Cu$_2$(C$_5$H$_{12}$N$_2$)$_2$Cl$_4$ since no plateau at half the
saturated value has been reported. With $J_1=2.4$ K, this means that $J_2'$ 
cannot
exceed 1.6 K. Although there is some discussion in the literature as to what the
actual value of this parameter is, all the estimates reported so far appear to
be smaller than this upper bound.

\section{\bf Conclusion}

In conclusion, we have shown that a strong coupling approach starting from the 
limit of strong rungs provides a simple and unifying picture of the very rich
physics that appears when ladders are put in a magnetic field. On one hand, it
gives a simple explanation of how
the Luttinger liquid physics emerges in the
intermediate phase of unfrustrated ladders. On the other
hand, this approach naturally leads to the presence of a plateau 
in the intermediate phase
at half the saturation value when the coupling between the rungs is strongly 
frustrated.
In systems where $J_\perp$ is effectively the largest coupling, this calculation
allows one to relate measurable quantities like $H_c^1$, $H_c^2$ and the 
critical exponents of the spin-spin correlation functions to the exchange 
integrals. Reported values for $H_c^1$ and $H_c^2$ in 
Cu$_2$(C$_5$H$_{12}$N$_2$)$_2$Cl$_4$ are well reproduced by this approach. It
will be interesting to analyze the critical exponents along the same lines when
experimental data are available. Finally, this approach provides quantitative
estimates of the frustration needed to create a plateau at half the
magnetization value for systems where the rung coupling is the largest one 
and should help in the search for systems exhibiting this
remarkable property.

I acknowledge useful discussions with C. Berthier, M. Horvatic, M.-H. Julien, L.
L\'evy and D. Poilblanc. I am especially indebted to Dr. A. Furusaki for
pointing out a mistake in an early version of this paper. 
After completion of this manuscript, I have been
informed by L. L\'evy that his group has independently obtained the same 
effective Hamiltonian in the case of unfrustrated ladders\cite{chaboussant4}.


\begin{figure}[hp]
\centerline{\psfig{figure=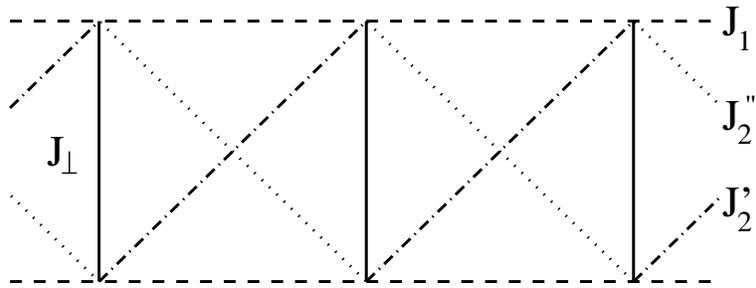,width=10.0cm,angle=0}}
\vspace{0.5cm}
\caption{Sketch of the ladder considered in this paper.}
\label{fig1}
\end{figure}


\end{document}